\documentclass[reprint,amsmath,amssymb,aps,nofootinbib]{revtex4}
\usepackage{graphicx} 
\usepackage{color,soul}
\usepackage{xcolor}
\newcommand{\be}{\begin{eqnarray}}
\newcommand{\ee}{\end{eqnarray}}

\newcommand{\tr}{\mathrm{Tr}}
\newcommand{\HH}{\mathcal{H}}
\newcommand{\rhot}{\tilde{\rho}}
\usepackage{braket}		
\usepackage{amsmath,amssymb}
\usepackage{palatino}
\usepackage{mathpazo}
\usepackage{graphicx}
\usepackage{mathrsfs}

\begin{document}

\title{\Large The Decoherent Arrow of Time and the Entanglement Past Hypothesis}

\author{Jim Al-Khalili\textsuperscript{1}}
\email{j.al-khalili@surrey.ac.uk}
\author{Eddy Keming Chen\textsuperscript{2}}
\email{eddykemingchen@ucsd.edu}

\affiliation{
\textsuperscript{1}School of Mathematics and Physics, University of Surrey, Guildford, GU2 7XH, UK
}

\affiliation{
\textsuperscript{2}Department of Philosophy, University of California, San Diego, 9500 Gilman Dr, La Jolla, CA 92093-0119
}

\date{\today}

\begin{abstract}
If an asymmetry in time does not arise from the fundamental dynamical laws of physics, it may be found in special boundary conditions. The argument normally goes that since thermodynamic entropy in the past is lower than in the future according to the Second Law of Thermodynamics, then tracing this back to the time around the Big Bang means the universe must have started off in a state of very low thermodynamic entropy: the \emph{Thermodynamic Past Hypothesis}. In this paper, we consider another boundary condition that plays a similar role, but for the decoherent arrow of time, i.e. the quantum state of the universe is more mixed in the future than in the past. According to what we call the \emph{Entanglement Past Hypothesis}, the initial quantum state of the universe had very low entanglement entropy. We clarify the content of the Entanglement Past Hypothesis, compare it with the Thermodynamic Past Hypothesis, and identify some challenges and open questions for future research. 

\end{abstract}

\maketitle

\section{Introduction}

In quantum mechanics, we appeal to decoherence as a process that explains the emergence of a quasi-classical order. 
Decoherence has no classical counterpart. Moreover, it is an apparently irreversible process 
\cite{Zeh1970,Zeh1971,Zeh1973,Leggett1980,Zurek1981,Zurek1982a,JoosZeh1985}.
In this paper, we investigate the nature and origin of its irreversibility. 

Decoherence and quantum entanglement are two physical phenomena that tend to go together. The former relies on the latter, but the reverse is not true. One can imagine a simple bipartite system in which two \textit{microscopic} subsystems are initially unentangled and become entangled at the end of the interaction. Decoherence does not occur, since neither system is \textit{macroscopic}.  Nevertheless, we will still need to quantify entanglement in order to describe the arrow of time associated with decoherence, because it occurs when microscopic systems become increasingly entangled with the degrees of freedom in their macroscopic environments. To do this we need to define entanglement entropy in terms of the sum of the von Neumann entropies of the subsystems.

\section{A Simple Model}

Consider a quantum system of interest, $S$, described by a pure state $\ket{\psi_S}$ at $t_0$, which is in a superposition of states in some basis $\{\ket{\chi_i}\}$. That is, say, $\ket{\psi_S}= \ket{\chi_1} + \ket{\chi_2}$. The system is surrounded by, but initially uncoupled from, a macroscopic environment in the state $\ket{E_0}$ at $t_0$. Once the system couples to this environment it will become entangled with it:
\be
 \ket{\psi_S}\ket{E_0}= ( \ket{\chi_1} + \ket{\chi_2})\ket{E_0}\ \ \rightarrow \ \ \ket{\chi_1}\ket{E_1} + \ket{\chi_2}\ket{E_2} = \ket{\Psi_{SE}},\ \ \ \ \ 
\ee
where $\ket{\Psi_{SE}}$ is the combined entangled pure state, and $\ket{E_1}$ and $\ket{E_2}$ are the ``pointer states'' that register the outcomes ``1'' and ``2'' in the macroscopic environment. The combined entangled state will obey unitary and time-reversal-invariant Schr\"odinger dynamics. However, given the entanglement, we can now no longer describe $S$ alone with a state vector. The density matrix of the combined system+environment is
\be 
\hat{\rho}_{SE} = \ket{\Psi_{SE}}\bra{\Psi_{SE}} = \sum_{i,j}^{2}\ket{\chi_i}\bra{\chi_j}\otimes \ket{E_i} \bra{E_j} ,
\ee
which means we can  describe the system of interest, $S$, with a reduced density matrix by tracing over the environment
\be
\tilde{\rho}_S &=& \mathrm{Tr}_E \bigg(\sum_{i,j}^2 \ket{\chi_i}\bra{\chi_j}\otimes \ket{E_i}\bra{E_j}\bigg)\nonumber\\
&=&\sum_{i,j}^2 \ket{\chi_i}\bra{\chi_j} \bra{E_j} E_i\rangle\ .
\ee
The off-diagonal matrix elements $\bra{E_j} E_i\rangle$ (for $i\neq j$), which decay over time, are the defining feature of decoherence, which takes place at the same time that  $S$ and $E$ become increasingly entangled (Eq.(1)). How quickly this takes place depends on how macroscopically distinguishable the states of the environment, $\ket{E_1}$ and $\ket{E_2}$, are. In realistic situations, decoherence is extremely efficient. In the scattering example studied by Joos and Zeh \cite{JoosZeh1985}, where a dust grain of size $10^{-3}$cm gets entangled with more and more air molecules in the environment (at normal pressure), the decay takes merely $10^{-31}$ seconds to suppress spatial interferences from the off-diagonal terms. 

This increasing entanglement of the system with its macroscopic environment and the inevitable decoherence of its reduced density matrix that results if the states of the environment are macroscopically distinguishable, is practically irreversible and gives us an arrow of time. Call this the \textit{decoherent arrow of time}. The huge number of degrees of freedom in the macroscopic environment means that this entanglement and inevitable decoherence are practically impossible to undo.\footnote{Two remarks here: (1) One might initially think that decoherence, as a system becomes increasing correlated (entangled) with its environment, is the same process as thermodynamic dissipation and the loss of heat energy of a warm body to its colder surroundings in classical thermodynamics. In some areas of physics this might be a useful association to make, but they are not the same. There can be an increase in entanglement between system and environment without any loss of energy at all. Decoherence does not require the environment to disturb the system and indeed happens on a much shorter timescale than any dissipation or relaxation. It can easily be shown that the rate of decoherence, $r(t)=\bra{E_1} E_2\rangle$, scales exponentially with the size of the environment. (2) There are cases of ``virtual'' decoherence that can be undone in controlled experiment, such as the reversible Stern-Gerlach experiment that combines the out-going beams \cite{Zeh1}. We shall focus on ``real'' decoherence \cite{schloss, zeh-p102}.   }

As pointed out by Gell-Mann and Hartle \cite{GH1994}, there are two ways to understand decoherence. The first is the decay (towards the future direction) of the off-diagonal terms in the reduced density matrix of the subsystem, as used in the example above as well as in derivations of master equations for decoherence \cite{schloss}. The second, mostly used in decoherence-histories approaches, quantifies how quasi-classical histories, described by series of projection operators that correspond to some macrostates, emerge from the  universal quantum state and evolve almost independently towards the future. The two definitions are deeply connected, but in our view they bring out different physical features of decoherence. The first depends on a partition of the combined system into a subsystem and its environment (or a factorization of the Hilbert space into tensor products) \cite{Zurek1998}. The second depends on a suitably chosen set of coarse-graining variables that correspond to the quasi-classical histories (or a decomposition of the Hilbert space into orthogonal subspaces). We will comment on both features in \S4 as they are connected to the decoherent arrow.    

The existence of a decoherent arrow of time is uncontroversial. But why it exists and how it connects with other arrows of time are questions worth exploring. Since the combined system plus environment is governed by the unitary Schr\"odinger equation, which is time-reversal invariant, then surely the interaction between them should be compatible with recoherence, and the un-doing of the entanglement between the system and the environment. Where then does the decoherent arrow of time come from, and how does it connect with other arrows of time? 

The puzzle regarding the first issue, the origin of the decoherent arrow of time, is similar to the long-standing problem in the foundations of thermodynamics regarding the origin of the thermodynamic arrow of time. The latter arrow is puzzling only if we focus on the time-reversal-invariant dynamical laws. However, if we postulate an initial condition that severely restricts the space of dynamical possibilities to those starting with a low thermodynamic entropy, the puzzle mostly goes away.  Questions about the origin of the thermodynamic arrow can be reduced to questions about the special initial condition, now called the Past Hypothesis \cite{albert}. We will refer to this initial condition as the {\em Thermodynamic Past Hypothesis} (TPH).

What is often done in discussions about decoherence and derivations of master equations of decoherence is a stipulation about the initial state: that it is a product wave function of the system and its environment -- or equivalently, a tensor product of their uncorrelated density matrices \cite{schloss, zeh}. Such an initial condition is of course not a logical requirement of the unitary Schr\"odinger equation, as the latter is compatible with other initial conditions. By following a similar logic as above, we are led to consider the cosmological version of this assumption \cite{zeh-p201}, according to which the universal wave function started off in a state of low entanglement entropy, as a tentative answer to the origin of product states in the universe and the decoherent arrow of time. 

In this paper, we first clarify the notion of entanglement entropy appropriate for describing the decoherent arrow, and discuss how it differs from other notions of entropy. We then formulate an \textit{Entanglement Past Hypothesis} (EPH) and explain how it depends on a partition of the universe into subsystems (or equivalently, a factorization of the Hilbert space of the universe into tensor products). Such a dependence differs from that on coarse-graining; it raises many questions about the status of EPH and the kind of explanation it provides. Finally, we discuss some connections to existing ideas and directions for future research.

\section{Notions of Entropy}

To get a clearer picture of the decoherent arrow, we need to be more precise regarding the appropriate notion of entropy. In this section, we discuss four different notions of entropy.\footnote{This is still far from an exhaustive list.} We suggest that the increase in entanglement entropy is the right measure for the decoherent arrow, because it satisfies two conditions: (1) it can be interpreted as an objective property of the universe, and (2) it is directly linked to decoherence. (Many ideas in this section are based on \cite{gold}, a useful survey of different notions of entropy in classical and quantum physics.) 

For a macroscopic isolated system of $N$ classical particles in a box, we can use the increase in Boltzmann entropy, $S_B$, to describe the relevant thermodynamic arrow of time as $\partial S_B/\partial t \ge 0$. 
The microstate of the $N$ particle system is a point in a $6N$-dimensional phase space and its macrostate is the set of phase points that are macroscopically indistinguishable, i.e. similar in values with respect to thermodynamic quantities such as temperature, pressure, and volume of the gas in the box. These macrostates, then, partition the phase space into regions. However, they are not all created equal, as some macrostates are overwhelmingly larger than others. The largest of them is the macrostate corresponding to thermodynamic equilibrium, which takes up almost all the volume in phase space. The classical Boltzmann entropy is proportional to the volume of the macrostate that includes the microstate of the system. More precisely, for a classical $N$-particle system whose microstate is $X$, with $X\in M$, where $M$ is the macrostate of the system, its classical Boltzmann entropy is:
\begin{equation}
    S_B (X) = k_B \text{log} |M|
\end{equation}
where $|\cdot|$ is given by the volume measure in phase space. Since the overwhelming majority of volume is taken up by the equilibrium macrostate, we expect a typical system starting in a low-entropy macrostate to rapidly move into the equilibrium macrostate. And given a postulate (the TPH, to be discussed in \S3) about the initial condition of the universe -- that it started in a low-entropy state -- we have an arrow of time for typical systems.     

We can generalize this picture from classical to quantum statistical mechanics.  For a macroscopic quantum system described by a wave function $\psi$ of gas in a box, we can use the increase in the so-called quantum Boltzmann entropy, $S_{qB}$ \cite{griffiths94,lebowitz08,goldstein10b}, to describe the relevant arrow of time. The microstate of the system is now a vector in a high-dimensional Hilbert space, $\HH$, rather than classical phase space, and its macrostate is a subspace of Hilbert space with wave functions that are macroscopically similar, i.e. similar in values of their thermodynamic properties. The macrostates, then, decompose the Hilbert space into orthogonal subspaces. However, again the macrostates are not created equal, as some are overwhelmingly larger than others, measured in terms of their dimensions. Just as in the classical thermodynamics case, the largest of them is the macrostate corresponding to thermodynamic equilibrium, whose dimension is almost equal to that of the full Hilbert space. The quantum Boltzmann entropy is proportional to the dimension of the macrostate, $\HH_M$, that includes the wave function of the system. More precisely, for a quantum system whose microstate is $\psi$, with $\psi$ in  $\HH_M$, the subspace corresponding to the macrostate of the system, its quantum Boltzmann entropy, is:
\begin{equation}
    S_{qB} (\psi) = k_B \text{log}\ dim (\HH_M)
\end{equation}
where $dim$ measures the dimension of the subspace. Since the overwhelming majority of dimensions is taken up by the equilibrium macrostate, we expect a typical system starting in a low-entropy macrostate to rapidly move into the equilibrium macrostate. And given a postulate about the initial condition -- the universal wave function has to start in a low-entropy state -- we have an arrow of time for typical universes. 

However, although we expect $S_{qB}$ to increase in time for realistic systems, it is not the right measure for the decoherent arrow of time. As many have long realized (for example see \cite{gold}), it is not the defining feature of the decoherent arrow. First, we can have a decoherent arrow without an increase of $S_{qB}$. Consider a product state that is already in thermodynamic equilibrium. We expect the components of the product state to become more entangled over time, displaying a decoherent arrow. However, the macrostate is already at its maximum value with respect to quantum Boltzmann entropy and will not increase over time. This relies on the important distinction between two different processes: dissipation (relaxation), which involves an exchange of energy between system and environment and which can take place both classically and quantum mechanically, and decoherence, which is a purely quantum effect.

This can also be seen in the Caldeira-Leggett master equation \cite{CL}, which is based on the simple model of a quantum system interacting with an infinite heat bath of harmonic oscillators and is derived using the path integral formalism of Feynman and Vernon \cite{feynman}. While the Caldeira-Leggett Master equation has its shortcomings (it relies on a Markovian approximation, weak coupling between system and environment and does not preserve the positivity of the reduced density operator), it nevertheless serves as a useful model to show the distinction between dissipation and decoherence. It is often written as
\begin{equation}
    \frac{\partial\rhot}{\partial t} = -\frac{i}{\hbar}\big[\hat{H},\rhot\big] -\gamma(x-y)\bigg(\frac{\partial\rhot}{\partial x} -\frac{\partial\rhot}{\partial y}\bigg)-\frac{2M\gamma k_BT}{\hbar^2}(x-y)^2\tilde{\rho}\ ,
\end{equation}
where $\gamma$ is the relaxation rate and $T$ is the temperature of the bath. The first term alone on the right hand side would give the von Neumann equation and describes the unitary Schr\"odinger dynamics. The second term causes dissipation: the loss of energy and a decrease of the average momentum, while the third term is responsible the decoherence (dephasing) term.

We can see here that the scenario described earlier in which the dissipation timescale is shorter than the decoherence timescale{\footnote{Ordinarily the decoherence timescale is many orders of magnitude shorter than the dissipation timescale.}} -- meaning that thermal equilibrium is reached due to dissipation {\em before} maximum entanglement -- corresponds to the situation in which the second term in the master equation dominates over the third term. This is of course an unlikely and unrealistic scenario since it requires strong coupling between system and environment and low temperature -- two conditions that are at odds with the assumptions made to derive the C-L master equation in the first place.

The opposite situation is also possible (and indeed more likely in most cases) whereby we can have an increase of $S_{qB}$ without a corresponding decoherent arrow. Consider a maximally entangled state starting in a macrostate of small quantum Boltzmann entropy. We expect it to evolve into larger and larger macrostates, corresponding to an increase in quantum Boltzmann entropy. But since it is completely decohered, it will not be more decohered in the future. This corresponds to the situation in which the third term on the right hand side of the master equation is much larger than the second term.

It is often suggested that the von Neumann entropy is a measure of the decoherent arrow. The problem, however, is that von Neumann entropy is stationary under the unitary dynamics for a closed system (such as that of the entire universe, which is after all the only truly closed system). It is analogous to the Gibbs entropy of a closed classical system. For a quantum system with quantum state $\rho$, its von Neumann entropy is defined as 
\begin{equation}
    S_{vN} = -\tr (\rho\log\rho)\ .
\end{equation}

When $\rho$ is a pure state, its von Neumann entropy is exactly zero and will remain at zero all the time it stays in a pure state isolated from its surroundings. This is certainly the case when we are talking about the quantum state of the universe, so its von Neumann entropy is not a sensible quantity to provide us with an objective arrow of time.

 At this point, one may try to use a mixed state to represent our ignorance of the underlying pure state of the universe, and use its von Neumann entropy as a measure of the decoherent arrow. However, insofar as the decoherent arrow is an objective property of the universe, it should be accompanied by an objective growth of some quantity that can be defined irrespective of how much information we have about the universe. 
 We therefore need to look for another measure of entropy.

Finally, we come to entanglement entropy, which we suggest is an appropriate measure of the decoherent arrow of time. 
Consider for simplicity a pure bipartite quantum system, $S$, partitioned into subsystems $A$ and $B$. Its total von Neumann entropy is zero.
Suppose the reduced density matrices for each of its subsystems are $\rhot_{A}=\tr_{B}(\rhot_{S})$ and $\rhot_{B}=\tr_{A}(\rhot_{S})$, where $\tr_{X}$ means a trace over subsystem $X$. We can define the entanglement entropy of the bipartite system $S$, with respect to the partition of $S$ into subsystems $A$ and $B$, to be the sum of $S_{vN}(A)$ and $S_{vN}(B)$: 
\begin{equation}
    S_{\text{ent}} (S) =     S_{vN}(\rhot_A) +  S_{vN}(\rhot_B) = -\tr(\rhot_A\log\rhot_A)  -\tr(\rhot_B\log\rhot_B).
\end{equation}
 The von Neumann entropy of $S$,  $S_{vN} (S)$, is always zero. However,  its entanglement entropy $S_{\text{ent}} (S)$ can increase, since the sum of $S_{vN}(A)$ and $S_{vN}(B)$ will increase if $A$ and $B$ are mixed states and continue to get more entangled with each other.

The definition of entanglement entropy depends on the partition being used.  For example, different ways of dividing the universe into subsystems can yield different values for the entanglement entropy. Each partition corresponds to a possible factorization of the Hilbert space of the system. The trivial partition of $S$ as a one-member set will of course yield an entanglement entropy of exactly zero. But now consider a pure state system that is partitioned as follows: 
\begin{equation}\label{example}
    \ket{\psi}_S = \frac{1}{\sqrt{2}} \big(\ket{0}_A \ket{1}_B + \ket{1}_A\ket{0}_B\big)\ket{0}_C
\end{equation}
where subsystems $A$ and $B$ are fully entangled together, with system $C$ treated as a separate subsystem. If the entangled state of $A$ and $B$ are treated as a single (pure) subsystem then, relative to this partition, which corresponds to the factorization of the Hilbert space of $S$ into the tensor product of $\mathcal{H}_{AB} \otimes \mathcal{H}_C$, it is a separable state whose entanglement entropy is zero. However, if we partition the system into the three subsystems $A$, $B$, and $C$, or factorize the Hilbert space as $\mathcal{H}_{A} \otimes \mathcal{H}_{B} \otimes \mathcal{H}_C$, then there is non-trivial entanglement among the subsystems, and the entanglement entropy of $S$ is non-zero.  Clearly, entanglement entropy is a well-defined quantity only relative to a particular choice of partition. Similarly, when we say a system is in an unentangled or product state, it is a meaningful statement only relative to some factorization of the Hilbert space of the system. 

In the previous example, two maps on the same Hilbert space yield two tensor products. On only one of them is the entanglement entropy of the global quantum state zero.
 
For this reason, to define entanglement entropy for the universal quantum state, $\Psi$, we should keep track of the partition that divides the universe into subsystems. Suppose the universe is divided into $N$ subsystems whose reduced density matrices are $\rhot_1, \rhot_2$,... and $\rhot_N$. The partition corresponds to a factorization of the Hilbert space of the universe, $F: \mathcal{H} \rightarrow \mathcal{H}_1 \otimes \mathcal{H}_2 \otimes ... \otimes \mathcal{H}_N$.  Then the entanglement entropy of the universal quantum state $\Psi$, relative to  partition $F$, is:

\begin{equation}
    S_{\text{ent}}^F (\Psi) = -\sum_{i,1}^N \tr(\rhot_i \text{log} \rhot_i)\ .
\end{equation}

Note that the number of partitions, $N$, is arbitrary and can be anything from 2 to $\infty$. This definition of entanglement entropy of the universal quantum state meets the two criteria for a good measure of the decoherent arrow. It is directly linked to decoherence. When the universe becomes more decohered, it corresponds to the situation when the subsystems become more entangled, and the entanglement entropy of the universal quantum state increases. It \textit{can} be interpreted as an objective property of the universe, relative to a partition. If the universe comes equipped with a fundamental structure of division into subsystems, the partition will correspond to a fundamental feature of the universe. However, if the partition is not fundamental, it may introduce some arbitrariness, which we discuss in the next section and suggest how it impacts the formulation of the initial condition about low entanglement entropy. There are other related measures of entanglement that can be used here, so $S_{\text{ent}}^F$ is not the only choice. But we shall focus on $S_{\text{ent}}^F$ as a particularly simple definition.

\section{The Entanglement Past Hypothesis}

We have suggested that entanglement entropy is a good measure for the decoherent arrow. However, there is still the question why there is a decoherent arrow in the first place. After all, the fundamental Schr\"odinger dynamics for the universe is time-reversal invariant. Recoherence, the process by which the entanglement between system and environment is undone, is also compatible with the dynamics. What then explains the fact that the quantum state in the past has much lower entanglement entropy than the current quantum state, which in turn has much lower entanglement entropy than future ones? 

That question has an analogue in the foundation of classical statistical mechanics. Given the time-reversal-invariant dynamics of classical mechanics, what explains the fact that the past has lower thermodynamic entropy than the future? A standard answer is to appeal to a special initial condition, the TPH, according to which the initial state of the universe is one with very low thermodynamic entropy (understood as Boltzmann entropy in statistical mechanics). Given the TPH, it is plausible to expect that almost all trajectories satisfying the constraint will display an entropy gradient. The assumption preserves the time-reversal invariance of the dynamical laws, but breaks the time-translation invariance by postulating this `special state' of the universe at $t=0$ (the Big Bang).  Nevertheless, it is regarded by many cosmologists as part of the best explanation for the thermodynamic arrow of time in our universe. Let us state it as follows:

\begin{description}
\item[Thermodynamic Past Hypothesis (TPH)] At one temporal boundary, the universe has very low thermodynamic entropy. 
\end{description}

One might understand the thermodynamic entropy in terms of classical or quantum Boltzmann entropy. In the case of a quantum universe, it should be understood as the requirement that the initial macrostate, represented by the Hilbert space subspace $dim(\HH_{PH})$, is small in dimension, so that the quantum Boltzmann entropy of the initial wave function of the universe, $S_{qB} (\Psi_0) = k_B \text{log}\, dim(\HH_{PH})$, is low. 

We propose a similar initial condition to explain the decoherent arrow: 
\begin{description}
\item[Entanglement Past Hypothesis (EPH)] At one temporal boundary, the universe has very low entanglement. 
\end{description}
By imposing EPH at one temporal boundary but not the other, it also breaks time-symmetry. This formulation of EPH is ambiguous, since it does not specify the measure of entanglement nor the factorization relative to which the entanglement is low.\footnote{It is actually close to some common statements about the origin of the decoherent arrow. Sometimes people posit that the initial quantum state is an unentangled state without saying which factorization of the Hilbert space is used and what justifies the choice \cite{zeh-p201, Albrecht2022}.} It is true that in practice (e.g. laboratory settings) we often know the system of interest and its environment, but here we are asking a further question, concerning which partition should be used when formulating the initial condition of the universal wave function.

Given the tools we have developed so far, we can clarify EPH as follows. We shall consider  some more precise versions of EPH and discuss their pros and cons. Suppose, as before, the universe is divided into $N$ subsystems whose reduced density matrices are $\rhot_1, \rhot_2$,... and $\rhot_N$, corresponding to a factorization of the Hilbert space of the universe, $F: \mathcal{H} \rightarrow \bigotimes^N_{i=1} \HH_i$. Relative to $F$, the entanglement entropy of the initial universal wave function has a particular low value, $m$. 
\begin{description}
\item[Entanglement Past Hypothesis$_m$ (EPH$_m$)] $S_{\text{ent}}^F (\Psi_0) = m$. 
\end{description}

EPH$_m$ is formulated with an explicit reference to a factorization of the universal Hilbert space. If there is a fundamental and precise principle for how to divide the universe into subsystems, it can be used as the basis for privileging a corresponding factorization in Hilbert space. However, in the absence of such a fundamental principle, the factorization dependence can appear problematic.  This is especially the case if EPH has the status of a fundamental law in the theory. After all, EPH is required to explain the decoherent arrow of time, but it is logically independent of the dynamical laws in quantum theory. (Its negation is compatible with the Schr\"odinger equation.) We should strive for a relatively simple and natural formulation of EPH, devoid of unnecessary arbitrariness. But the particular factorization $F$, which EPH$_m$ refers to, would be arbitrary; there seems to be no fundamental rule about how we should divide the universe into subsystems and no smallest size below which we cannot divide them further.  

Recall that TPH is also dependent on a ``partition,'' but it is (in the quantum case) an orthogonal decomposition of the universal Hilbert space into subspaces $\HH\rightarrow \bigoplus^K_{i=1} \HH_i$. The subspaces $\{\HH_i\}$ are selected based on the choice of some thermodynamic variables, such as energy, temperature, and pressure. One might argue for such a choice based on considerations of robust regularities at a higher level \cite{albert}. However, the way EPH$_m$ depends on the partition  is different, as it corresponds to a specific way of carving the universe into subsystems, subregions in space, or subsets of degrees of freedom. It is hard to see how one might argue for a particular partition of physical space or a factorization of Hilbert space. 

Moreover, in addition to the dependence on a partition, a complete understanding of the decoherent arrow of time will require the decoherent histories approach (\S2), which depends on some choices of coarse-graining variables, represented by sets of orthogonal projectors that sum to unity in the Hilbert space of the universe. So the appeal to EPH does not eliminate the decomposition-dependence, but it adds a further factorization-dependence. In this sense,  EPH may be more problematic than TPH.

 Zurek, a pioneer and proponent of the decoherence program, fully recognizes \cite{Zurek1998} the factorization dependence as an important foundational issue for understanding decoherence: 
\begin{quote}
    In particular, one issue which has often been taken for granted is looming large, as a foundation of the whole decoherence programme. It is the question of what the `systems' which play such a crucial role in all the discussions of the emergent classicality are.
\end{quote}

Some progress has been made in this direction \cite{cotler2019, bao}, but it may be fruitful to explore whether we can formulate a precise version of EPH without picking a particular factorization.

Let us consider some potential solutions that avoid the dependence on a particular partition. As a first attempt, we postulate that $S_{\text{ent}}^F (\Psi_0)$ has a low value, $m$, \textit{on any factorization $F$ that maps the universal Hilbert space into tensor products}. However, on pain of inconsistency, we cannot fix a precise value for $m$, unless it is zero. In general when $F_i$ and $F_j$ are two different factorizations of the universal Hilbert space, if $S_{\text{ent}}^{F_i} (\Psi_0)=m$, then $S_{\text{ent}}^{F_j} (\Psi_0) \neq m$. (See (\ref{example}) for an example.)

To implement this idea, and to avoid the inconsistency arising from different factorizations, we may require that $S_{\text{ent}}^{F_i} (\Psi_0)$ is zero on every factorization:
\begin{description}
\item[Entanglement Past Hypothesis$_{0}$ (EPH$_0$)] $S_{\text{ent}}^{F_i} (\Psi_0) = 0$ on every factorization $F_i: \mathcal{H} \rightarrow \bigotimes^N_{i=1} \HH_i$. 
\end{description}
EPH$_0$ is very strong, placing a severe constraint on the initial state. It is a precise and factorization-independent statement of the requirement that the initial quantum state be absolutely unentangled. If it is true, EPH$_0$ can be highly informative. However, there are two ways it could be too strong. First, there may be no vector in the universal Hilbert space compatible with EPH$_0$.  Second, there may be other considerations that require the initial state to be not absolutely unentangled. 

Let us consider two ways to relax the requirement imposed by EPH$_0$. We can require that $S_{\text{ent}}^{F_i} (\Psi_0) = 0$ on every factorization $F_i$ of a certain class:
\begin{description}
\item[Entanglement Past Hypothesis$_{0R}$ (EPH$_{0R}$)] $S_{\text{ent}}^{F_i} (\Psi_0) = 0$ on every factorization $F_i$ of a certain class $R$. 
\end{description}
As an example, we may consider only factorizations that correspond to partitioning the universe into \textit{spatial} regions.  We can also require that it is not exactly zero, but only less than or equal to some small number $m$. 
\begin{description}
\item[Entanglement Past Hypothesis$_{\leq m}$ (EPH$_{\leq m}$)] $S_{\text{ent}}^{F_i} (\Psi_0) \leq m$ on every factorization $F_i$. 
\end{description}
This is consistent with the fact that different factorizations can yield different values of the entanglement entropy, as it just requires that the different values arising from different partitions are bounded from above by $m$. We may also combine the two ways to weaken EPH$_0$ as follows:
\begin{description}
\item[Entanglement Past Hypothesis$_{\leq mR}$ (EPH$_{\leq mR}$)] $S_{\text{ent}}^{F_i} (\Psi_0) \leq m$ on every factorization $F_i$ of a certain class $R$. 
\end{description}

Both EPH$_0$ and EPH$_{\leq m}$ take away the undesirable dependence on a particular factorization of the universal Hilbert space and thus a particular way of splitting the universe into subsystems. EPH$_{0R}$ and EPH$_{\leq mR}$ are not entirely factorization-invariant, but they do not depend on any particular factorization as they quantify over all factorizations of a certain class $R$, to be specified as part of the theory.

\section{Interpretations and Further Questions}

In this section, we discuss three issues related to the EPH: its status, connections to other ideas, and questions for future research. 

First, when we postulate a special initial condition, we are confronted with a question regarding its status in the physical theory. Is it purely an accidental feature of our universe unexplained by any fundamental laws of physics, or can it be derived from the laws? 

For the EPH, the answer to this question depends on how we view the decoherent arrow of time. Should we understand it as merely an accidental feature of our universe, or a lawful regularity? We offer some considerations that it is the latter. Decoherence and the temporal asymmetry it encodes are crucial for understanding how quantum mechanics works and how the quasi-classical order emerges in a quantum universe. More concretely, it helps to explain the apparent collapse of the wave function in experimental situations. In a quantum experiment, when we measure the state of the physical system, it undergoes an apparent collapse into one of the definite states. The temporal reverse of this is a recohering process. If EPH is not assumed, then different components of the wave function can recohere and give rise to a ready state of the measurement system, erasing past records of measurement results. But this would make it impossible to trust our records about past measurements, undermining the empirical support for quantum mechanics. To avoid such undermining, we should postulate EPH as an additional law constraining the feature that the initial state of the universe must have.\footnote{There is a more radical approach. We may view EPH as something completely perspectival or subjective, as arising from our subjective perspective. This will have the consequence that anything grounded in EPH, such as the decoherent arrow of time, will become entirely perspectival and subjective. }


Should EPH be regarded as a fundamental law, or can it be derived from other fundamental laws we already postulate? It cannot be derived from the fundamental dynamical laws alone since the Schr\"odinger equation is compatible with the negation of EPH. The two are logically independent. 

Can EPH be derived from other boundary condition laws? There are two cases here. The first case concerns boundary condition laws such as TPH, which selects a set of wave functions, one of which is supposed to be the correct one. If we already regard TPH as a fundamental law, perhaps we can derive EPH by adding a probability distribution compatible with TPH. If we impose a uniform probability distribution $\mu$ over wave functions compatible with TPH, perhaps we can expect that $\mu-$most wave functions satisfying the constraint of TPH will also satisfy EPH. However, this is not true, because most wave functions will be highly entangled. Unentangled wave functions are atypical on the standard measure. Nevertheless, if $\mu-$most wave functions compatible with TPH will become more entangled and decohered in the future, that will also be sufficient to explain the decoherent arrow of time. In that case, we will no longer need to postulate EPH. 

The second case concerns boundary condition laws that choose a particular and precise initial state. It is possible that EPH can be derived from such postulates. For example, the Hartle-Hawking No Boundary wave function \cite{hartle}, a particular choice of the initial condition, either validates EPH or fails EPH. If it validates EPH, then EPH would be redundant in the Hartle-Hawking theory. If it fails EPH, a proponent of the Hartle-Hawking theory needs another way to explain the decoherent arrow of time. 


One final speculative idea worth exploring briefly here is whether the TPH is, if not derivable from the EPH, at least a result of the EPH. This is in the opposite direction from the earlier idea of deriving EPH from TPH. Assuming the universe began very small it would have been at or near maximum thermodynamic entropy (very hot and dense but near thermodynamic equilibrium). As it expanded and cooled the light elements began to form before the density and temperature dropped below the threshold needed for further nuclear fusion, thus trapping regions of low entropy. Later, gravitational clumping allowed for the fusion to continue in stars, which could radiate out this low entropy energy to drive the universe we see today \cite{WallaceGravity, RovelliEntropy}. Thus, it is the rapid expansion of the universe that allowed for these low entropy regions to appear while allowing for the universe as a whole to increase in thermodynamic entropy as more phase space became available. The point is that the expansion of the universe is ultimately responsible for the thermodynamic arrow of time. We have restated the above scenario in order to suggest extending the story back to an earlier cause: namely what might have caused the expansion of spacetime in the first place?

One bold suggestion is that spacetime might itself emerge due to quantum entanglement at the Planck scale. This is not a new idea and has been suggested by others. For example, in describing the early universe in the language of quantum field theory in curved spacetime, Bao {\em et al} \cite{bao} suggest that spacetime is ultimately made up of co-moving regions (spacetime modes), each described by a finite number of quantum degrees of freedom. These emerge from ``a store of zero-energy Planck scale spacetime modes initially in their vacuum states, which become dynamical when their wavelengths grow longer than the Planck length.''

At the start, all these spacetime modes are unentangled, but once a few become entangled with each other, forming the initial spacetime structure, their increasing entanglement with surrounding vacuum modes brings more into the fold and spacetime expands as the entanglement spreads.

So, whether one argues that it is entanglement that drives the expansion of the universe or that the expansion allows for increasing entanglement, either way, we need an EPH to start the universe off.

Given the discussion in \S4 and this section, we make two final observations. First, requiring EPH to be a fundamental law in effect requires EPH to be a simple statement, since we expect fundamental physical laws to be simple and compelling \cite{chensimplicity}. That precludes versions of EPH that depend on a particular or arbitrary choice of factorization, which can be extremely complicated to specify. For example, a particular factorization can correspond to a specific partition of physical space into some small (but non-zero) regions, and to describe the partition cells one by one is a complicated business. In contrast, EPH$_{0R}$ and EPH$_{\leq mR}$ can be specified more simply, since $R$ can correspond to the class of all partitions of physical space into small regions with cell size $r^3$, for some number $r$. There is less difficulty accepting EPH$_{0R}$ and EPH$_{\leq mR}$ as simple candidate laws.

Second, there is some tension between the requirement that EPH should be simple and hence (as much as possible) factorization invariant and the idea that spacetime is emergent from a quantum description (as in the final bold suggestion). Both EPH$_{\leq 0R}$ and EPH$_{\leq mR}$, when $R$ corresponds to some class of spatial factorizations, still depends on a fundamental spatial structure. The idea that minimal quantum entanglement is partly defined in terms of spatial structure is in conflict with the idea that spatial structure is not fundamental but derived from quantum entanglement. That is an apparent circularity that needs to be resolved. The observation is a concrete payoff of the previous discussion, where we tried to be more precise in the formulation of EPH. It may point to a general difficulty in simultaneously satisfying the requirement of invariance and the program of obtaining space(time) from something more fundamental.

\section{Conclusion}

We suggest that the decoherent arrow of time can be understood as an objective feature of the universe, arising from a special initial condition that we call the Entanglement Past Hypothesis (EPH). To make sense of EPH, we are led to similar considerations about the thermodynamic arrow of time, but we find the two arrows to be conceptually distinct. Whether they can be unified is a question we leave to future work. The standard formulation of EPH depends on a factorization of the Hilbert space of the universe into tensor products. We explore some tentative solutions that take away such a dependence. Despite the simple setting and open questions it raises, we hope the discussion is useful for highlighting the possibilities and the challenges in understanding the nature and the origin of the decoherent arrow of time.  

\section*{Acknowledgements}
The authors would like to thank Emily Adlam, David Albert, Jeff Barrett, Dorje Brody, Craig Callender, Sean Carroll, Paul Davies, Sheldon Goldstein, Andrea Rocco, Simon Saunders, Karim Thebault,  David Wallace,  Ken Wharton, and the audience at the 2023 UC San Diego Workshop on Arrows of Time for useful discussions. 
This work was made possible thanks to the support of the John Templeton Foundation through grant number 62210.


\end{document}